\documentclass[manuscript,screen]{acmart}
\usepackage{booktabs} 
\usepackage[ruled]{algorithm2e}
\usepackage{array}
\usepackage{color}

\newcolumntype{P}[1]{>{\centering\arraybackslash}p{#1}}
\SetAlFnt{\small}
\SetAlCapFnt{\small}
\SetAlCapNameFnt{\small}
\SetAlCapHSkip{0pt}
\IncMargin{-\parindent}

\usepackage[ruled]{algorithm2e} 

\acmJournal{CIE}
\acmVolume{9}
\acmNumber{4}
\acmArticle{39}
\acmYear{2010}
\acmMonth{3}

\setcopyright{acmcopyright}

\acmDOI{0000001.0000001}

\begin{document}

\title{Measuring third party tracker power across web and mobile} 
 \titlenote{}
 \subtitle{}
 \subtitlenote{}

\author{Reuben Binns, Jun Zhao, Max Van Kleek, Nigel Shadbolt}
\orcid{0000-0003-4718-2190}
\affiliation{%
  \institution{University of Oxford, Department of Computer Science}
  \streetaddress{Wolfson Building, Parks Road}
  \city{Oxford}
  \postcode{OX1 3QD}
  \country{UK}}
\email{reuben.binns,jun.zhao,max.van.kleek,nigel.shadbolt@cs.ox.ac.uk}
\renewcommand\shortauthors{Binns, R. et al}

\begin{abstract}


Third-party networks collect vast amounts of data about users via web sites and mobile applications. Consolidations among tracker companies can significantly increase their individual tracking capabilities, prompting scrutiny by competition regulators. Traditional measures of market share, based on revenue or sales, fail to represent the tracking capability of a tracker, especially if it spans both web and mobile. This paper proposes a new approach to measure the concentration of tracking capability, based on the reach of a tracker on popular websites and apps. Our results reveal that tracker prominence and parent-subsidiary relationships have significant impact on accurately measuring concentration.

\end{abstract}

%
%


%
%




\maketitle

\section{Introduction}

\begin{quote}
``Power in the digital economy is partly driven by the degree to which a given undertaking can actually, potentially or hypothetically collect and diffuse personal information." ~\cite{hustinx2014privacy}
\end{quote}

Billions of people use digital devices every day, generating vast amounts of data. The range of devices available to consumers proliferates, from gaming consoles to smart televisions, but perhaps the most significant source of data about consumer behaviour are websites and smartphone applications. In what has been described as `the Internet's original sin', the monetisation of this data became a primary business model of the current digital era ~\cite{zuckerman2014internet}. Firms with the ability to collect such data have become a significant part of the economy~\cite{acquisti2016economics}, with the online advertising industry earning \$59.6 billion per year in the U.S. alone \cite{iab2016}.

A large share of this revenue is accrued by `third-party' trackers ~\cite{montes2015value}; companies who gain access to user data as a result of their technology being integrated by a `first-party' entity into its website or mobile application. Third-party tracking potentially raises greater privacy concerns than first-party data collection because it can often link records from multiple websites or apps to a single user. This can provide a more complete picture of an individual, from where they shop, to their social networks and likely political opinions. The more websites and apps that choose to integrate a particular third party tracker, the greater that tracker's power to collect personal information from a wider range of sources and build more comprehensive profiles.



The accumulation of power in the hands of third-party tracker networks has not escaped the attention of regulators. While much of this attention has come from privacy and data protection regulators, competition authorities also have begun facing calls to consider potential impacts arising from the consolidation of tracking capability in a single firm or small number of firms \cite{pasquale2013privacy,harbourdissenting,lande2008microsoft,geradin2013}. 

However, the market power of trackers is not obviously encapsulated by traditional measures used by competition and antitrust regulators, which focus on a firm's `market share'; usually defined in terms of its revenue or unit sales as a percentage of the industry total. These measures make sense when the focus is on preventing monopolies and oligopolies from forming and charging higher prices. However, the power of third-party tracking technology arises from its capacity to gather, analyse and monetise personal data from a wide array of sources in multifarious ways; this may not directly correspond to the sales and revenues accrued by the firm in any given year. As a result, traditional measures fail to reflect relevant dimensions of the power that can potentially accrue to firms operating third-party tracker networks.

Measuring market concentration also involves defining the scope of the relevant market. This raises an important challenge in the context of tracking, because it can operate across different digital ecosystems. The World Wide Web and smartphones provide potentially quite different environments in which tracking can occur.\footnote{Other platforms, such as smart televisions, or Internet-of-Things devices, may also develop their own particular tracking ecosystems. In what follows we focus exclusively on web and mobile as they are the most significant platforms at present, but acknowledge that third-party tracking through other devices may soon become more prevalent} Each platform has a different set of risks and opportunities; mobile application developers are constrained by the app marketplace providers and OS-level privacy tools, while website publishers have to reckon with the protective measures put in place by browsers ~\cite{anderson2010inglorious}. Some tracking firms may therefore focus exclusively on native mobile applications, others on the web, while others may operate across both. Previous large-scale studies of tracking have tended to focus on either the web (e.g. ~\cite{englehardt2016census} ) or mobile (e.g. ~\cite{vallina2016tracking} ), without linking the observed tracking technologies to the firms that own them and may be operating counterparts on other platforms. The precise extent of overlap between firms operating trackers on each platform is therefore unknown, but highly relevant to any assessment of their market power.

There is therefore a need for reliable data on the extent to which third-party tracking technology is integrated on these two large platforms, which can be linked to the firms behind such technology, in order to inform our definition of the tracking market. Furthermore, we require a consistent, alternative metric to meaningfully measure the market share of a tracker firm, to truthfully reflect its `tracking capability' in the market. In the absence of such data and metrics, regulators lack any rigorous means of assessing the potential threats arising from the accrual of tracking capability in the hands of firms operating third-party trackers.


In this paper, we aim to address these challenges by outlining a novel combination of data collection and integration methods, and a corresponding metric for assessing a firm's share of tracking capability, which can better inform regulators' activities.


\subsection{Contributions}
The main contributions of this paper are as follows:

\begin{enumerate}

\item We propose a novel metric for measuring the concentration of power within the tracking ecosystem(s), called PROWISH-HHI~(PROminence Weighted Integration Share - Herfindahl-Hirschman Index), which incorporates not only the extent to which a firm has its tracking technology integrated into first-party services, but also its prominence and parent-subsidiary relationships.


\item We identify the crossover between the web tracking and mobile tracking ecosystems, by comparing the distribution of trackers on 5,000 websites and 5,000 Android applications, as well as a direct comparison of the trackers present on 358 services which have both web and mobile versions. We identify that the overlap between the tracker ecosystems associated with these two platforms is partial.

\item We find that the web and mobile tracking ecosystems are similarly competitive, although a somewhat different set of firms dominate on each platform. Interestingly, a prevalence-based, subsidiary-level analysis gives the appearance that the web tracking ecosystem is highly competitive; but when prominence and parent-subsidiary relationships are taken into account, it transpires that it is just as concentrated as the mobile tracking ecosystem.

\item Using data from our knowledge base of tracker firms, we model the impact of mergers and acquisitions on the market concentration of each ecosystem. We find several cases from recent years which have likely increased market concentration levels, one of which is well in excess of the threshold at which a regulator would typically intervene.

\end{enumerate}


\section{Background}

This section further outlines the motivation for our analysis. We begin by defining some essential technical and economic aspects of third-party tracking. We then outline how consolidation of tracking capability is an increasingly important consideration in competition and antitrust policy; why existing measures of market concentration are inadequate; and why new metrics based on alternative data are needed.

\subsection{Third-party tracking}
Third party tracking can be broadly defined as any transfer of personally identifying data from an online service, to an entity other than the provider of that service. For example, a website provided by a news organisation (the `first party') might include code from a targeted advertising network (the `third party'). When a user visits the website, the third party code may set a cookie on the user's browser, so that when the user visits other websites which have also integrated the same third party tracker's code, their activity can be linked from across these sites. Alternatively, they may `fingerprint' the user by recording some uniquely identifying attribute or combination of attributes such as their browser settings or installed plugins.

Tracking can be used for multiple purposes, including targeted advertising and marketing and analytics. Within the targeted advertising industry, there are typically multiple agents providing different services in a complex, multi-sided market. This includes not only advertisers and `publishers' (i.e. first party providers of websites and apps), but a range of intermediaries, including networks of third parties with tracking technology, intermediary data brokers, and exchanges which allow advertisers to compete against one another for targeted advertising opportunities via automated auctions.


More recently, regulators have argued that the consolidation of tracking capability and its effects on privacy should be considered in competition and antitrust activity ~\cite{schechner2017germany}. In a 2014 report, the European Data Protection Supervisor (EDPS) suggested that `if market power in the digital economy can be measured according to control of commercialisable personal information, then merger decisions could in turn take account of the market effects of combining these capabilities' ~\cite{hustinx2014privacy}.

\subsection{Measuring market power of third-party trackers}

If they are to incorporate considerations about tracking capability into competition policy, regulators require a systematic way of measuring this form of market power. Competition and antitrust regulators rely on such measures in order to determine, in a fair and systematic way, when to intervene in a market.

As noted in the EDPS report:
\begin{quote}
`Measuring control of personal information would be
a challenging exercise. A relevant market share held by a provider of a free online service cannot easily be calculated by reference to data on traditional sales or volume. These difficulties could be surmounted if ... authorities were to collaborate ... in developing a standard for measurement of market power in this area' ~\cite{hustinx2014privacy}
\end{quote}

The challenge is therefore to develop a metric for market share, which focuses on third-party tracking capability. This could be considered alongside more traditional measures such as revenue, sales or volume from digital advertising or other business models (where market concentration is already known to be high ~\cite{slefo2017inapp}). An alternative metric would provide additional information that could complement traditional measures in two ways: 

\subsubsection{Disconnection between tracker power and traditional measures of market share}
First, a firm's revenues may bear little direct relation to their tracking power. The ability to track users across multiple sites and apps can be monetised in different ways, from advertising to the provision of market insights. Tracking capability might be directly monetised by selling user data, or an essential part of a company's long-term path towards becoming profitable. The ability to turn a profit from operating a third-party tracking network may therefore not generally be directly proportional to the power inherent in it. As such, a firm with relatively poor tracking capability might post high revenues, while another with vast capability might not. Aside from revenues, another common way to measure a firm's market power is by share of unit sales. However, given the wide-ranging, heterogeneous and unusual ways in which third-party tracking can generate income, there is no clear standard `unit' of sale. Advertising impressions are just one aspect of some trackers' business models, and the third parties which serve advertisements may not be the ones undertaking the tracking used to target them. The closest candidate for this purpose may be the rates paid by advertising exchanges for user data (which have been estimated in \cite{olejnik2013selling}); but this is only one business model for monetising personal data.


\subsubsection{Market definition}
Second, there is the common problem of market definition. In the context of advertising, different platforms, such as print and TV, tend to be treated as different markets, such that a single firm's presence in one market would not be taken into account in assessing the legality of their making an acquisition in another.\footnote{See e.g. COMP/M.5932 News Corp/BSkyB [2010])} In the context of firms operating third party tracking networks, it is not clear whether they should be treated as one market or several different markets. One important consideration is whether such networks are generally confined to a single platform, i.e. the web or mobile. If there are a significant number of firms operating on both platforms, it may be more appropriate to consider them as a single market. While some companies are known to operate in both, the extent of overlap is currently unknown, because revenues for those companies may not be broken down by platform. In some cases, companies operating tracking networks may themselves not even be able to provide such a breakdown, for instance if they supply a platform-agnostic API which is used by developers to integrate into both mobile and web services. Regulators therefore need evidence about the distribution of tracking technology belonging to particular firms on both platforms in order to make judgements about market definition.

\section{Related Work}

The third party tracking ecosystem has been studied on both the web and mobile using a variety of methods. Large scale web studies have analysed the presence of third-party trackers by inspecting network traffic associated with a website. This data can be gathered either through crowd-sourcing (e.g. \cite{vallina2016tracking,yu2016tracking}) or with an automated web crawler (e.g. \cite{englehardt2016census}). These methods have provided evidence on the prevalence and distribution of trackers on the web, and the long tail distribution of web tracking. Several leading companies have been repeatedly observed and confirmed~\cite{roesner2012detecting,libert2015exposing,yu2016tracking}, each of which can capture more than 20\% of a user's browsing behavior. 

Several studies of third-party tracking have also been conducted on mobile platforms~\cite{vallina2016tracking,book2015empirical}, using both dynamic and static detection methods. Dynamic methods, as in web-based tracking studies, involve inspecting network traffic from the browser / device and identifying any third party destinations that relate to tracking. One common approach has been OS-level instrumentation, such as those of TaintDroid~\cite{enck2014taintdroid}, and AppTrace~\cite{qiu2015apptrace}.  An alternative to low-level OS instrumentation is to analyse all communications traffic transmitted by an app whilst it is in use~\cite{ren2016recon}. Other methods involve unpacking an application's source code (on Android systems, this comes as an Android application package(APK)) and detecting the usage of third-party tracking libraries~\cite{arzt2014flowdroid,batyuk2011using,egele2011pios,lin2014privacygrade}.

Other aspects of tracking have been studied, including the variety of techniques that are used, from cookies ~\cite{aziz2015cookie,englehardt2015cookies,englehardt2016census} to fingerprinting \cite{acar2013fpdetective}. A more recent field study by Yu et al provided a finer-grained view into tracker behaviour, by classifying data being transmitted to trackers as either `safe' or `unsafe' ~\cite{yu2016tracking}. Another factor is the permissions requested by an app, which constrain the kinds of data a third party can obtain; longitudinal research has found that Android apps request additional privacy-risking permissions on average every three months ~\cite{taylor2017}.

The crossover between the mobile and web tracking ecosystem has also attracted attention in recent research. As part of their study on the mobile tracking ecosystem, ~\cite{vallina2016tracking} also compared the list of mobile trackers with known web trackers published in existing blacklists, and confirmed a big gap between the mobile tracker ecosystem and the web one. The most closely related to work is research by~\cite{leung2016recon} who compare privacy risks between using a web version of a service and its mobile version. They compare 50 services between their web and mobile versions to identify the extent to which personal identifiable information (PII) is being transmitted on different platforms. However, while this provides insight into the different levels of sensitivity of the \emph{data} trackers get between platforms, it does not identify the distributions of different firms involved in the two tracking ecosystems at large scale. More broadly, such empirical work has generally not been informed by an understanding of the corporate relationships of tracking firms, or the impact of mergers and acquisitions on the overall market structure.



\section{Measuring trackers and tracker market}

Motivated by these challenges, we develop an alternative measure which aims to more directly capture the tracking capability of a third-party network relative to the overall tracking industry. It combines methods from empirical research on tracker networks in computer science, with traditional measures of market concentration in economics. In this section we provide an overview of the related research, motivations and rationale behind this alternative measure.

\subsection{Calculating tracker prominence}

As mentioned above, most previous studies report the `prevalence' of a tracker across a corpus of websites or apps, i.e., the number of first-parties it is integrated with.\footnote{E.g. ~\cite{roesner2012detecting,libert2015exposing,vallina2016tracking}} This is useful in ranking different trackers according to their popularity with developers. However, popularity with developers may not directly relate to the number of individuals whom a tracker may track, because some trackers might be placed on more popular websites or applications. This would give them the capacity to track more users. In order to capture this aspect, Englehardt et al \cite{englehardt2016census} measure the `prominence' of a tracker, defined as:

$$\texttt{Prominence}(t) = \sum_{edge(s,t)=1} \frac{1}{rank(s)}$$

where $edge(s,t)$ represents that third party tracker $t$ is present in site or app $s$.

The advantage of a prominence-weighted metric is that it takes into account the `reach' of a tracker, that is, the number of instances of installed applications through which it can track an individual. This is based on the assumption that rank is roughly proportional to a website's audience (actual user numbers would be more accurate, but are unfortunately unavailable). From the perspective of measuring market power, this may be more informative than a straightforward count of the number of websites or applications it has been integrated with. A tracker which is only integrated on websites or apps with very few users will impact fewer users.\footnote{However, being tracked on a niche website may be considered more compromising of privacy - see section 7.3 below for discussion of this point.}

\subsection{The Herfindahl-Hirschman Index}\label{sub:hhi}
Prevalence and prominence are useful as absolute measures of a tracker's reach. However, from an economic competition perspective, neither can tell us how much power a tracker has relative to other trackers. This is necessary if we want to measure the market concentration of the tracker industry, and use this as a basis for assessing the likely effects of consolidation on competition.

The Herfindahl Hirschman Index, or HHI, is a standard measure used by regulators to determine the level of concentration in a market. It is based on the following formula:

$$\texttt{HHI} = \sum\limits_{i=1}^N s_i^2$$

Where $s_i$ is the market share of a firm $i$ in the market, and $N$ is the number of firms.

In standard economic analyses of market concentration, an HHI below 0.01 indicates a highly competitive industry, below 0.15 indicates an unconcentrated industry, between 0.15 and 0.25 indicates moderate concentration, and above 0.25 indicates a highly concentrated industry ~\cite{warren1990implications,shapiro20102010}. These values are used by regulators to justify intervention in a market. In the U.S., for example, a merger or acquisition which raises the HHI above 0.25 would be sufficient for intervention. In the E.U., for any market which already shows a concentration of more than 0.1, consolidations which raise the HHI by more than 0.025 will be a cause for concern ~\cite{verouden2004role}. Market concentration measurement is therefore a key part of the operation of antitrust and competition regulation.

\subsection{Calculating tracker market share}

Market share is usually defined in terms of revenue or unit sales. For reasons discussed above, these measures may bear little relation to the extent to which a firm has the ability to track individuals across multiple websites and apps. Instead, this aspect of market share may be better captured by focusing on the first-parties which the tracker is integrated into. A simple way to capture the market share of a tracker would be to look at its \emph{integration share}:

\begin{definition} \label{def:share}
The \emph{integration share (ISH)} of a tracker firm $t_i$ is measured by its prevalence as a proportion of the sum prevalence of all trackers $t_j$in a given tracker market $T$. 
$$s_{t_i} = prevalence(t_i) / \sum\limits_{j=1}^N prevalence(t_j)$$
\end{definition}

However, as noted above, the prevalence of a tracker does not capture the number of people who are potentially subject to being tracked by it. This is because a tracker which is integrated into many unpopular websites / apps might reach less people than a tracker which is integrated into a few very popular websites / apps. We therefore propose an alternative measure of market share, based on the prominence of a tracker as a proportion of the sum total prominence of all trackers in the market. We call this the prominence-weighted integration share (PROWISH). The ISH and PROWISH are used below in conjunction with the HHI to measure the concentration of tracking power in the web and / or mobile tracking ecosystem(s).

\begin{definition} 
The \emph{prominence-weighted integration share (PROWISH) HHI} of a tracker firm $t_i$ is measured by its prominence, as a proportion of the sum prominence of all trackers $t_j$ in a given tracking market $T$. 
$$s_{t_i} = prominence(t_i) / \sum\limits_{j=1}^N prominence(t_j)$$
\label{def:p-share}
\end{definition}

We believe that these two metrics (ISH- and PROWISH-HHI) could usefully address the challenge of measuring tracker power. While the total share of sales and revenues from marketers is one important aspect of a competitive market, share of total tracking capacity as measured by ISH/PROWISH is a key input to business success in this sector. A high ISH/PROWISH-HHI might mean that smaller or new firms entering the tracking market may find it difficult to compete with dominant incumbent firms. This could be due to a relatively small number of first-party integrations (ISH) and / or the small combined userbase associated with those first-party integrations (PROWISH).

\subsection{The Relationship Between Tracker Market Concentration and Privacy Risk}

Before presenting the results of this study, it is worth considering how these measures of tracker market concentration (ISH/PROWISH-HHI) relate to privacy risks. On the one hand, concentration of tracking capacity could indicate negative consequences for consumer privacy, because a) the more websites or applications that a particular third-party tracker is integrated with, the more comprehensive and revealing the profiles it can build, and b) the available apps / websites consumers have to choose from are more likely to contain one or more of the dominant tracker companies.
This has led to calls for such risks to be addressed in competition and antitrust policy; if a firm attempts to significantly expand its share of tracking capacity, for instance through acquiring another firm, authorities should examine the potential impact on privacy ~\cite{swire2007protecting}. Mergers and acquisitions are a common occurrence in data-related sectors as a whole, with 55 deals in 2008 and 164 in 2012 ~\cite{oecd2015data}. Authorities tasked with vetting mergers and acquisitions between firms with access to large amounts of personal data have increasingly considered potential privacy harms arising from the consolidation of tracking capabilities ~\cite{geradin2013}.

However, the relationship between high concentration of tracking capacity and privacy risk is not always in the same direction. There are also ways in which a lower concentration could be bad for privacy, in so far as privacy preferences involve minimising the total number of firms that have the capacity to track a single user. For instance, if a new third party enters the market, assuming first parties integrate it without substituting any existing third parties, this would increase the number of trackers that users of those first parties are exposed to, whilst also reducing market concentration. Similarly, if two firms merge, this would increase market concentration whilst reducing the overall number of entities that (certain) users are exposed to.

Whether such concentration is better or worse for privacy may depend on consumer privacy preferences. In previous work we have found that users express two kinds of preferences in relation to trackers, which are sometimes in conflict ~\cite{van2017better}. One is to ~\emph{minimise the total number of distinct tracking entities they are exposed to}. The other is to ~\emph{minimise the total number of distinct ways in which they may be tracked by any single entity}; i.e. any given tracker should have access to data from the smallest number of distinct services as possible, because this prevents any one tracker from knowing too much. From a `minimise entities' perspective, a highly concentrated market might be favourable if there are just a few large entities responsible for all the tracking a user is exposed to. From a `minimise distinct leaks to any single tracker' perspective, a highly concentrated market may be be worse, because those few large entities are able to track users via multiple different services. We therefore cannot assume that a more concentrated tracker market will be worse for privacy, as it depends on the circumstances and the nature of individual privacy concerns.

\section{Data collection and analysis}
Our tracker concentration measurement is based on the detection of third party trackers in websites and smartphone applications. As we reviewed earlier, for both platforms, there are multiple ways to detect third parties. Furthermore, not all third parties are necessarily `trackers'. Tracker identification therefore involves two steps; first, detecting third party activity, and second, selecting only those third parties that are actually trackers (for some definition of tracker).

\subsection{Third party detection}
To identify third parties on the web, we chose to use and extend an existing web crawl dataset published by the OpenWPM project.\footnote{OpenWPM - https://github.com/citp/OpenWPM} Similar to many other crawling tools (such as FPDetective~\cite{acar2013fpdetective}), OpenWPM simulates real browsing activity using a headless browser, intercepting and recording traffic between the browser and remote servers. A typical OpenWPM dataset includes information about all HTTP requests and responses during the visit of a site, as well as any cookies transmitted or fingerprinting techniques applied, depending on the crawling set-up. 

The dataset we used as the basis of our study was Webcensus, part of the Web Transparency project, and consisted of a stateful (i.e., cookie-based) crawl of 100k web sites,~\footnote{http://webtransparency.cs.princeton.edu/webcensus/} downloaded in July 2016. After loading the Postgres data dump into a local data store, we use a Python script to identify third parties for each of the top 5,000 websites listed on Alexa (also retrieved in July 2016). Approximately 400 of these top Alexa sites were not present in the 100k Webcensus dataset, so we set up a local instance of OpenWPM to fill in this gap. To identify potential trackers, our script identified all third-party hosts, for each web site, by filtering all requests to first-party destinations.

To identify potential third party trackers on mobile applications, we used a combination of static and dynamic analysis. This analysis was based on 5,000 of the most popular Android applications in the Google Play Store. Apps were selected based on the number of downloads as reported on the marketplace, during the summer of 2016. The Android application packages for each app were downloaded directly from the Play Store via two Android devices (Nexus 7 and 6P), using automated Python scripts and the Monkey Runner API\footnote{https://developer.android.com/studio/test/monkeyrunner/MonkeyRunner.html}.

We began by performing dynamic analysis of mobile app traffic for a subset of 260 apps from our corpus of 5,000. This was conducted on a Nexus 7 device using a man-in-the-middle proxy,\footnote{mitmproxy.org} and a user session for each app was simulated. Network traffic from each session was collected and all hosts connected to during the session were logged.\footnote{Code for this analysis can be found at \emph{https://github.com/sociam/PROWISH}} First-party hosts associated with each app were filtered out at a later stage. Any new hosts were added to the corpus of hosts found in the website analysis. We also conducted a first round of static analysis of all 5000 apps, to identify third party libraries. We unpacked APKs using APKtool,\footnote{See https://ibotpeaches.github.io/Apktool/
} and examined their package structure. This revealed a wide array of third party libraries used by each app which were potentially facilitating tracking.\footnote{We evaluated several existing static code analysis tools. However, every existing tool seems to have a different focus on the type of libraries they want to detect, for example, libraries calls with implicit security risks, or detection of obfuscation techniques used in the code of an APK. In the end, these existing tools provided too low a recall for our data collection.}

\subsection{Identifying popular trackers amongst third party hosts and libraries}

Having found \textasciitilde 5000 third party hostnames and \textasciitilde 9000 third party libraries associated with the websites and applications, we then sought to identify popular trackers from amongst them. One approach to identifying trackers would be to use known \textit{tracking protection lists}, as in previous studies~\cite{englehardt2016census,yu2016tracking}.\footnote{E.g. Ghostery \url{https://www.ghostery.com}, Adblock \url{https://getadblock.com} or Disconnect \url{https://disconnect.me}} This method may optimise precision, since each entry on the list has been identified manually. However, these lists may not be fully representative of the trackers in existence. They may be oriented towards the particular circumstances of the organisations and user groups who compiled them (such as geography, sites of interest, etc.). As such, relying on them for classification might lead us to miss prevalent trackers. Another limitation of tracking protection lists is that they tend to be web-centric. Such lists are largely geared towards the tracking ecosystem on the web and have very low coverage for trackers in the mobile tracking ecosystem ~\cite{vallina2016tracking}. Finally, they may have different definitions of tracking; sometimes they are advertising-oriented rather than tracking in general, and some lists make exemptions for trackers who sign up to codes of conduct.\footnote{See e.g. AdBlock Plus' 'acceptable ads' scheme ~\cite{meier2014erfolgreicher}}

For these reasons, rather than using any existing blacklists, we use our own criteria to determine which third-parties are trackers. Our basic definition of \textbf{a third party tracker} is \emph{an entity that collects data about users from first-party websites and / or apps, in order to link such data together to build a profile about the user}. 

For instance, a behavioural advertising company whose code is deployed on a website or app, records user behaviour, and uses a unique identifier to aggregate this behaviour across different sites and / or apps, would be considered a tracker. By contrast, a content delivery network, integrated on multiple first party apps or websites, which did not record and link user behaviour, would not be considered a tracker. Since this definition focuses on the behaviour of third parties which may be taking place invisibly on a server, it is not always possible to determine whether an entity is a tracker just by looking at the data sent to it from the user device. Instead, we rely on publicly available information about the legal entity responsible for the activity taking place on the server associated with the hostname, or for the code in the library, to determine if it is a `tracker'. Sources included Crunchbase, OpenCorporates, and Wikipedia.

As we were only interested in the more significant trackers, we discarded any third party hostnames / libraries from the set of 14309 which had less than 0.5\% coverage in our mobile / web datasets. This yielded a set of 766 hostnames and 215 libraries to investigate further.

In the case of the libraries, we searched the web for further information about the function of the library and the entity responsible for it. In most cases the primary sources were developer documentation associated with the library. Using this information, 75 of the 215 libraries we investigated met our definition of tracking.\footnote{Some of these libraries were unofficial open source libraries written by external developers which allow, for instance, integration of behavioural advertising via an official API. For our purposes, these unofficial libraries were still associated with the company since their presence in the code indicates that the app is capable of facilitating user tracking by that tracker.} Many of the other libraries were related to core application code (e.g. parsers, data storage tools, game engines, or crypto toolkits). These were not considered to be trackers. Furthermore, for a small proportion ($< 3\%$) of the apps, it was found that obfuscation was applied to third party libraries, which resulted in scrambled identifiers and truncated package structures that made it difficult to identify the original identity of the library. These packages were omitted from our analysis. 

In the case of domains, we identified the organisation responsible for the domain. We used the WHOIS registration records, as well as other public information sources about the company, to assess whether the organisation likely engages in a tracking activity.\footnote{Public information resources included Crunchbase, Wikipedia and OpenCorporates} In cases where our search of public information could not conclusively determine whether the organisation engaged in tracking, we also examined the traffic logs corresponding to a putative tracker's host. We decoded the call payloads, and searched for identifiable information using a range of regular expressions designed to detect personal data types such as device ID, email address, or location.\footnote{Code for the dynamic analysis can be found at \url{https://github.com/sociam/xray/blob/xray-refine/mitm/}}

After tracing the company associated with each host or library, we were able to consolidate multiple different hosts / libraries under the same company name. In order to aid our market structure analysis, we created a knowledge base of information about tracker companies compiled from public sources. As well as mapping third-party libraries and hostnames to their associated companies, it also includes information about each company, including its jurisdiction, date founded, and any parent-subsidiary corporate structures they might be part of. This also allowed us to conduct additional analyses which collapse subsidiaries by their parent companies, thus avoiding double-counting of apparently distinct trackers on a website / app when they are owned by the same parent company.

Finally, with this knowledge base of tracker entities, and their associated hostnames and libraries, we were able to re-analyse our datasets of web traffic logs and decompiled APKs to measure the distribution of these trackers across apps and websites. The third party hostnames present in the WebCensus crawl were matched against known trackers in our knowledge base. In the case of the apps, we incorporated an additional level of static analysis, whereby the decompiled code (both the app itself and any third-party libraries) was recursively searched for strings containing HTTP/HTTPS URLs. Any such URLs that matched a known tracking domain were recorded. The library-based and url-based methods were then combined to yield a set of trackers for each app. We found that this combination of static methods yielded the highest recall, and captured the majority of trackers that were found through purely dynamic analysis.\footnote{See appendix A for further discussion of these methodological issues.}

\subsection{Identify web and mobile pairs}
It is possible that differences observed between the mobile and web tracking ecosystems based on the 5,000 web sites and apps might be due to differences in the kinds of services that are offered on each platform. For instance, the top 5,000 Play Store applications include many games, which rarely have any equivalent web versions, and therefore any gaming-oriented trackers might be more prevalent amongst apps in this tracking dataset. We therefore also compared two subsets of genuinely equivalent services, in order to explore this possibility.

These equivalent services were selected by first automatically searching for APK names corresponding to the top 5,000 website names, and vice-versa the top 5,000 applications. For each website, we queried the Google Play store with that website name and selected the first result. Similarly, for a package called \emph{com.example} (following the Java package naming conventions), the TLD was reversed and a request was made to the website example.com. This yielded \~900 possible equivalent website-app pairs. These were manually validated to find out whether the website and application were provided by the same source and offered the same service, resulting in 358 genuinely equivalent pairs. These pairs were subjected to the same analysis as the 10,000 websites and apps.

\section{Results}

We organise our results as follows. We begin by summarising the most prevalent and prominent trackers on both the web and mobile ecosystems. Then we present analysis of the similarity and difference between each platform's tracking ecosystem, using both our 5,000 websites and apps (as well as further validation of overlap rates between our 358 equivalent web-app pairs). Finally, we present a range of market concentration indices for each ecosystem, showing how the use of parent-subsidiary relationship data and prominence-weighting can significantly affect the overall HHI value.

\subsection{Which trackers are most prevalent and prominent on the web and mobile?}

Here we give an overview of the trackers we identified from the top 5,000 web sites and top 5,000 mobile apps from the Google Play Store in terms of their prevalence (the number of apps / websites they are integrated into) and prominence (where prevalence is weighted by the popularity of the app / website).


\paragraph{\textbf{Top prevalent trackers on the web}}

\begin{figure}[h!]
\centering
\begin{tabular}{cc}
  \includegraphics[width=62mm]{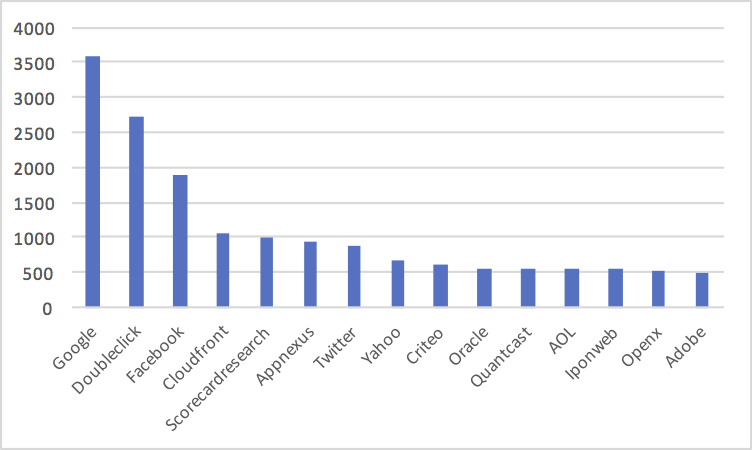} &   \includegraphics[width=62mm]{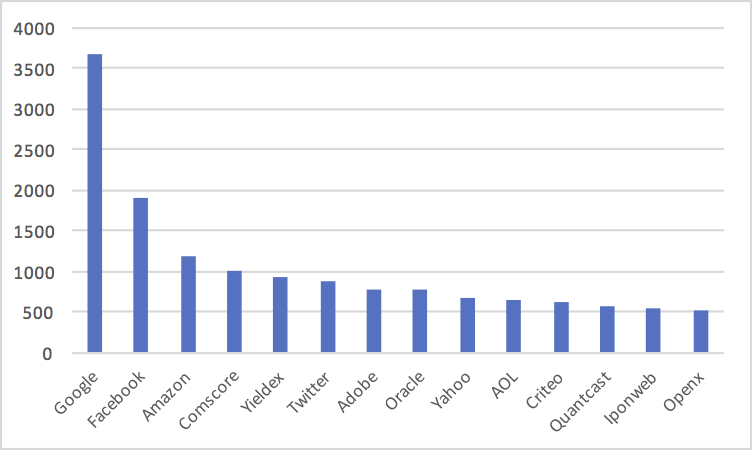} \\
(a) & (b)  \\[6pt]
\end{tabular}
\caption{Trackers that are present in over 500 web sites (subsidiaries (a), and parent firms (b)).}
\label{fig:webtracker}
\end{figure}

In the most popular 5,000 web sites (according to Alexa in July 2016), the total number of trackers that are present in at least two first party web sites is just over 2,000, among which 253 are present in over 25 sites. This confirms the long tail observation from previous studies~\cite{yu2016tracking}. Furthermore, our data also shows that tracking on the web is dominated by the top 15 trackers, each of which accounts for at least 10\% of the tracking on the 5k web sites (see Figure~\ref{fig:webtracker} (a)).

Using our knowledge base of the subsidiary relationships between tracker companies (for example, \textit{DoubleClick} as part of \textit{Google / Alphabet}), we observe a very different set of leading tracker companies, see Figure~\ref{fig:webtracker} (b). Now \textit{Amazon}, \textit{comScore}, \textit{Yieldex}, and \textit{Adobe} also become leading tracking companies because of the multiple smaller subsidiary tracker companies that they own. For example, Adobe owns Demdex, Omniture and LiveFyre with a combined prevalence of 740.

\paragraph{\textbf{Top prominent trackers on the web}}
When weighting the presence of these trackers by the \emph{popularity} of the websites on which they appear, i.e., using the prominence metric outlined before, we observe quite a different ranking for many trackers (Table~\ref{table:change}). For instance, \textit{Yieldex}, despite being the 5th most prevalent web tracker, was only the 11th most prominent, meaning that the websites it is integrated with are on average less popular than those websites which similarly prevalent trackers are integrated with. 

When comparing the overall rank change for web trackers from their prevalence to prominence values, we find that about 30\% of web trackers become less prominent even though they have a good reach, which indicates that they are not always present in the most popular web sites. However, the prominence measurement also reveals some powerful web trackers that were less known when they were measured by their reach. For example, the ranking of \textit{Taobao} has moved up 125 places, even though it is only present in 42 web sites. 

\begin{table}[h!] 
\caption{Rank differences between prevalence and prominence of web trackers\label{table:change}}{
\parbox{.47\linewidth}{
\centering
\small
\begin{tabular}{P{0.3in}|c|P{0.4in}|P{0.45in}}\\
\textbf{Prev. rank} &  \textbf{Tracker} & \textbf{Prev.} & \textbf{Rank change} \\\hline
1. & Google         & 3676  & 0 \\
2. & Facebook     & 1898  & 0\\
3. & Amazon       & 1187  & -3 \\
4. & Comscore    & 999   & -1\\
5. & Yieldex        & 927    & -6 \\
6. & Twitter         & 865    & -2 \\
7. & Adobe         & 767    & 0 \\
8. & Oracle         & 763    & +4 \\
9. & Yahoo         & 662  & -1 \\
10. & AOL         & 642  & +7 \\
\end{tabular}
}
\hfill
\parbox{.47\linewidth}{
\centering
\small
\begin{tabular}{P{0.3in}|c|P{0.7in}|P{0.45in}}\\
 \textbf{Prom. rank} & \textbf{Tracker} & \textbf{Prom.} & \textbf{Rank change} \\\hline
1. & Google	& 19.30744847 & 0 \\
2. & Facebook	& 3.805494974 & 0 \\
3. & AOL		& 1.619862177 & +7 \\
4. & Oracle	& 1.473848421 & +4 \\
5. & Comscore	& 1.283621696 & -1 \\
6. & Amazon	& 1.252199961 & -3 \\
7. & Adobe	& 1.176659952 & 0 \\
8. & Twitter	& 1.169940505 & -2 \\
9. & Alibaba	& 0.978651143 & +77 \\
10. & Yahoo	& 0.97482352 & -1  \\
\end{tabular}
}
}
\end{table}


\paragraph{\textbf{Top prevalent trackers on mobile}}

\begin{figure}[h!]
\centering
\begin{tabular}{cc}
  \includegraphics[width=62mm]{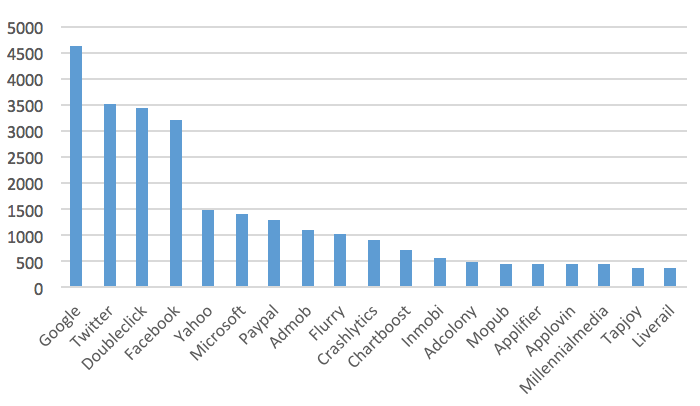} &   
  \includegraphics[width=62mm]{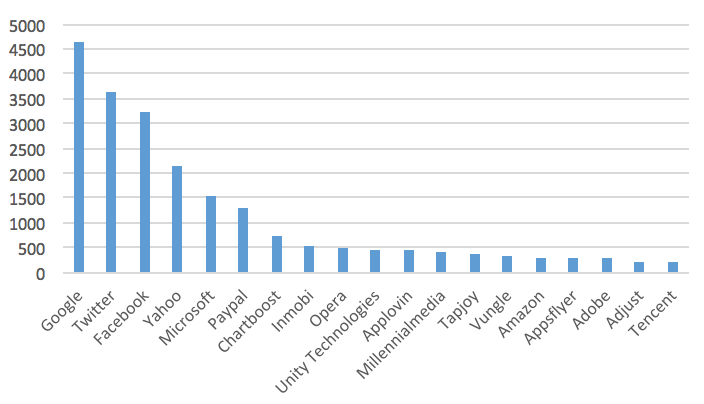} \\
(a) & (b)  \\[6pt]
\end{tabular}
\caption{Top 20 trackers that are present in over 100 mobile applications (subsidiaries (a), and parent firms (b)).}
\label{fig:mobtracker}
\end{figure}

We found a similar long tail distribution of trackers in the mobile ecoystem, although only 83 mobile trackers met our threshold of 0.5\% ecosystem coverage, compared to 253 web trackers. After aggregating mobile trackers by their parent companies, we again found that the ranking of tracker companies changed. 
 (see Figure~\ref{fig:mobtracker}).


\paragraph{\textbf{Top prominent trackers on mobile}}
As with web trackers, we also calculated the \emph{prominence} of the mobile trackers based on the popularity of the applications on which they appear. The difference between prevalence and prominence rankings were less pronounced than for the web, with the same firms occupying similar places in the top 10 (Table~\ref{prev_prom_app}). 

\begin{table}
\caption{Rank differences between prevalence and prominence of mobile trackers\label{prev_prom_app}}{
\parbox{.47\linewidth}{
\centering
\small
\begin{tabular}{P{0.25in}|P{0.75in}|P{0.35in}|P{0.4in}}\\
\textbf{Prev. rank} &  \textbf{Tracker} & \textbf{Prev.} & \textbf{Prev. - Prom. rank} \\\hline
1. & Google	& 4635 & 0 \\
2. & Twitter &	3630 & 0 \\
3. & Facebook &	3218 & 0 \\
4. & Yahoo	& 2124 & -1 \\
5. & Microsoft &	1522 & 1 \\
6. & Paypal	& 1286 & 0 \\
7. & Chartboost	& 718 & 0 \\
8. & Inmobi	& 553 & -1 \\
9. & Opera	& 500 & 1 \\
10. & Unity Technologies &	450 & 0 \\
\end{tabular}
}
\hfill
\parbox{.47\linewidth}{
\centering
\small
\begin{tabular}{P{0.25in}|P{0.7in}|P{0.35in}|P{0.4in}}\\
 \textbf{Prom. rank} & \textbf{Tracker} & \textbf{Prom.} & \textbf{Prom. - Prev. rank} \\\hline
1. & Google & 7.19 & 0 \\ 
2. & Twitter & 5.918 & 0 \\ 
3. & Facebook & 5.184 & 0 \\ 
4. & Microsoft & 2.596 & -1 \\ 
5. & Yahoo & 2.594 & 1 \\ 
6. & Paypal & 1.509 & 0 \\ 
7. & Chartboost & 0.849 & 0 \\ 
8. & Opera & 0.797 & -1 \\ 
9. & Inmobi & 0.793 & 1 \\ 
10. & Unity Technologies & 0.720 & 0            \\ 
\end{tabular}
}
}
\end{table}

\subsection{\textbf{How much crossover is there between mobile and web tracking ecosystems?}}
While some common names appear in both the top mobile and web tracker rankings, there are also some differences. To explore these commonalities and differences further, we compared the top web and mobile trackers from the 5,000 web sites and 5,000 apps to find out the degree of overlap between them. 

Of those trackers with more than 25 instances, only 33 subsidiary-level and 26 parent-level trackers were found in both the mobile and web ecosystems (see Table~\ref{table:overlap}). Some of these trackers occupy similar positions in each ecosystem, for example, \textit{Google} and \textit{Facebook} are first and third in both respectively. However, some trackers are clearly much more prevalent in the mobile ecosystem, such as \textit{Tencent} which was the \#26 most common mobile tracker but only \#253 in the web ecosystem.

\begin{table}[h!]
\caption{Trackers (parent firms) appearing in both the 5,000 web sites and 5,000 mobile apps\label{table:overlap}}{
\centering
\scalebox{0.9}{
\begin{tabular}{c|c|P{1in}|P{1in}}\\
\textbf{Tracker} & \textbf{Web rank} & \textbf{Mobile rank} & \textbf{Web - mobile rank} \\\hline
\textbf{Google} & 1 & 1 & 0 \\
\textbf{Twitter} & 2 & 6 & -4 \\
\textbf{Facebook} & 3 & 3 & 0 \\
\textbf{Microsoft} & 5 & 26 & -21 \\
\textbf{Paypal} & 6 & 185 & -179 \\
\textbf{Amazon} & 15 & 2 & 13 \\
\textbf{Adobe} & 17 & 7 & 10 \\
\textbf{Tencent} & 19 & 240 & -221 \\
\textbf{Alibaba} & 20 & 84 & -64 \\
\textbf{Comscore} & 25 & 4 & 21 \\
\textbf{Vk} & 27 & 110 & -83 \\
\textbf{Yandex} & 28 & 49 & -21 \\
\textbf{Akamai} & 30 & 53 & -23 \\
\textbf{Mail.Ru} & 31 & 121 & -90 \\
\textbf{Mixpanel} & 32 & 95 & -63 \\
\textbf{AOL} & 40 & 13 & 27 \\
\textbf{Spearhead} & 42 & 160 & -118 \\
\textbf{Moatads} & 43 & 24 & 19 \\
\textbf{Cloudflare} & 44 & 17 & 27 \\
\textbf{Rubicon} & 46 & 16 & 30 \\
\textbf{Opsmatic} & 54 & 21 & 33 \\
\textbf{Smartadserver} & 55 & 58 & -3 \\
\textbf{Signal} & 61 & 140 & -79 \\
\textbf{Smart Adserver} & 64 & 127 & -63 \\
\textbf{Pinterest} & 66 & 109 & -43 \\
\textbf{I-Mobile} & 72 & 227 & -155
\end{tabular}}
}
\end{table}


We conducted the same analysis on the 358 services (see Section 5.4), which are available in both an app and website format, in order to see if the degree of overlap in trackers differs when an app and website are versions of the same service. We found again that, apart from \textit{Google}, \textit{Facebook}, and \textit{Twitter}, which were in the top 10 for both platforms, no other trackers had significant presence on both platforms for this subset of 358 equivalent app-mobile pairs.


In order to further examine the degree of overlap between equivalent web and mobile versions of services, we analysed the rate of overlapping trackers, as shown below:

$$ overlapping\_rate_{app_i} = \frac{|mobile\_tracker_{app_i} \cap web\_tracker_{app_i}|}{|mobile\_tracker_{app_i} \cup web\_tracker_{app_i}|} $$

Using the the same combination of static app analysis, and dynamic website analysis, we found that the average app-website pair had 15\% of app-associated trackers and web-associated trackers in common. We also conducted an overlap analysis of the same pairs using the dynamic method described above, for both platforms, in case there were certain trackers undetected by our static method.\footnote{In this case, we simulated the website activity using the same mobile device (Nexus 7) with the standard Android browser.} This yielded a slightly higher overlap rate of 17\% (see appendix for further discussion of these methods). Whether 15\% or 17\%, this small proportion of overlapping trackers suggests that service providers generally choose to integrate different trackers depending on the platform they are developing for. Examples where trackers serving similar purposes differed between web and mobile versions include a popular book-reading service which appears to use Appsflyer for re-targeting on mobile but Criteo for similar purposes on its web version; and an online marketplace which appears to use Apteligent and Adjust for analytics on the mobile app but not on the web version.



\subsection{How concentrated is the tracker industry?}
Having collected the baseline prevalence and prominence figures for the top trackers on top web and mobile applications, we calculate the market concentration of these tracker ecosystems individually and combined.

As explained in Section~\ref{sub:hhi}, we use the HHI to calculate market concentration, which requires a measurement of each tracker's market share. A simple metric for measuring the market share of a tracker is its integration share (see~Definition 4.1). The alternative prominence-weighted integration share (PROWISH) may provide a more appropriate measure as it takes into account the popularity of an app or website (defined in Definition 4.2). Here the HHI is reported (see Table~\ref{table:marketstructure}) for both measures (denoted as \emph{ISH-HHI} and \emph{PROWISH-HHI}) for comparison.

Based on the tracking data of the 5,000 web sites, the web tracking ecosystem has an ISH-HHI of 0.02, suggesting a highly competitive industry according to U.S. Federal Trade Commission merger guidelines~\cite{shapiro20102010}. This value is almost exactly the same whether we measure it at the subsidiary or parent level. However, when the market share of a tracker is weighted by the prominence of the websites on which it appears, the market concentration is higher; PROWISH-HHI is 0.06 at the subsidiary level, and doubles to 0.12 at the parent level. In other words, using the PROWISH metric, combined with parent-subsidiary mapping, we find that the web tracking market concentration level is 6 times higher than when using the simple integration share metric (and would be sufficient to merit attention under standard E.U. competition measures).

By contrast, the mobile tracking ecosystem for the 5,000 apps has an ISH-HHI of 0.06 at the subsidiary level and 0.09 at the parent company level. When weighted by prominence, the ecosystem shows a concentration of 0.12 (subsidiary) and 0.08 (parent level). Interestingly, in contrast to web tracking, prominence-weighted parent-level measurement resulted in lower concententration. This seems in large part due to the fact that certain highly prominent subsidiaries very often co-occurr on websites / apps (e.g. Google and Doubleclick); consolidating them into one thus reduces the overall prominence of the parent company and the HHI in turn.

Since our analysis also suggests that the mobile and web tracking ecosystems are to some degree overlapping, we also calculated their combined market concentration, by merging both datasets and summing the PROWISH values of firms operating across both platforms. This combined web and mobile tracker PROWISH-HHI value would be relevant if a regulator were to decide to treat the two tracker ecosystems as a single market. Merging the two ecosystems in this way resulted in concentration levels somewhere between the levels found under each platform individually.

\begin{table}[h!]
\caption{HHI for web and mobile tracker ecosystems computed at different levels\label{table:marketstructure}}{
\centering
\scalebox{0.9}{
\begin{tabular}{lll}
\textbf{Market} & \textbf{ISH-HHI} & \textbf{PROWISH-HHI} \\\hline
Web trackers (subsidiary) & 0.02 & 0.065 \\
Web trackers (parent) & 0.02 & 0.12 \\
Mobile trackers (subsidiary) & 0.06 & 0.10 \\
Mobile trackers (parent) & 0.09 & 0.12 \\
Web and mobile trackers combined (subsidiary) & 0.032 & 0.037		\\
Web and mobile trackers combined (parent) & 0.037 & 0.047        
\end{tabular}}
}
\end{table}

The disparity in market concentration observed between the subsidiary and parent-level shows the importance of mapping corporate relationships, and demonstrates that mergers and acquisitions can have important impacts on market structure. In order to explore this further, we modelled the effect of particular acquisitions that have taken place over recent years. We calculated the delta between the \emph{actual} PROWISH-HHI and what it would be if the subsidiary company (or companies) were not owned by their parents (i.e. if they had remained independent and grown at the same rate) (Table~\ref{table:acquisitions}). In many cases there was no difference, because the parent company has no significant presence in the tracker ecosystem aside from that of its subsidiary or subsidiaries. However, in other cases, the PROWISH that parents gained from subsidiaries accounted for significant impacts on the overall HHI, suggesting that such acquisitions merit attention from regulators. The most significant case is the merger between Google and Doubeclick (highlighted in bold), which accounts for a difference in PROWISH-HHI of 0.039 and would, if happening now, would meet the proposed threshold under EU law.\footnote{In modelling the effects of mergers and acquisitions, we do not have access to historical data about how large parents and subsidiaries were when they merged / were acquired. Our reported figures are therefore based on the hypothetical effect of trackers merging given their current distributions.} 

\begin{table}[h!]
\caption{Effect of acquisitions on market structure.\label{table:acquisitions}}{
\centering
\scalebox{0.9}{
\begin{tabular}{llll}
\textbf{Parent} & \textbf{Acquisition(s)}          & \textbf{Platform} & \textbf{PROWISH-HHI difference} \\\hline
\textbf{Google}          & \textbf{Doubleclick} (2007)               & Web               & 0.039                           \\
Twitter         & Mopub (2010), Crashlytics (2011) & Mobile            & 0.006                           \\
Adobe           & Demdex (2011), Lyvefire (2016)   & Web               & 0.004                           \\
Aol             & Adap.tv (2013)                   & Web               & 0.004                           \\
Alibaba         & Umeng (2013)                     & Web               & 0.003                           \\
Microsoft       & Linkedin (2016)                  & Web               & 0.001                           \\
Facebook        & Liverail (2014)                  & Web               & 0.0004     
\end{tabular}}
}
\end{table}

\section{Discussion and Conclusions}

\subsection{Insights into measuring tracker market structure}
Our findings suggest third-party tracking on the web and mobile is widespread and diverse. A small number of firms dominate, and a long tail of less significant trackers exists on both platforms. These findings are broadly in accordance with those from previous studies of web and mobile trackers. However, by combining prominence measures, studying the web and mobile in tandem, factoring in parent-subsidiary knowledge, and adapting traditional economic measures of market concentration, we can derive a number of novel insights.

First, our findings demonstrate the importance of mapping hostnames and libraries to companies, and incorporating parent-subsidiary relationships into analysis of trackers. The ranking of trackers (both in terms of prevalence and prominence) changed significantly when they were aggregated by parent companies. Working at the level of legal entities, rather than at the level of hostnames or library names, enables inferences to be drawn about the structure of the market. Furthermore, by accounting for the popularity of the websites a tracker appears on, our PROWISH metric is sensitive to the number of users each tracker is likely to affect.

Second, measurements of the structure of a tracker market are highly sensitive to the nuances of prominence and parent-subsidiary relationships. This is particularly clear when we consider that the HHI for web tracking is just 0.02 at the subsidiary-level, but rises six-fold to 0.12 when prominence and company ownerships are taken into account. This demonstrates that prevalence values of a particular tracker may not be sufficient to assess its market power. Competition authorities should therefore use empirical data about tracker distributions with caution, and take into consideration tracker prominence and parent-subsidiary relationships.

Third, our findings indicate that there is no neat division between mobile-oriented trackers and web-oriented ones. Almost half of the popular trackers on mobile are also popular on the web, but those that are well-spread on mobile are not all well-spread on the web and the firms that dominate each ecoystem are not all the same. There are many firms which appear to be focused on only one platform; but a substantial number operate in both. Of those which operate across platforms, just a handful occupy similar positions (only Google, Twitter and Facebook have a top 10 rank in both). This means that the tricky question of market definition cannot be easily resolved and may have to proceed on a case-by-case basis. It is hard to draw firm conclusions about why certain services choose to use different third party trackers for their web and mobile versions, but possible reasons include that they may outsource development of web and mobile applications to different agencies; some trackers may be easier to integrate on a given platform; and mobile / web versions may be developed at different times, betwen which the cost-benefit analysis of integrating a particular tracker might have changed.

Finally, the PROWISH-HHI values for web and mobile were the same at the parent-level, suggesting that both platforms support relatively competitive tracking industries. However, it is worth noting that the web tracker industry appears to be less oligopolistic than mobile; the top 8 web trackers have a combined prominence share of 54\%, compared to 65\% for the top 8 mobile trackers. For comparison, this is a higher concentration ratio than for real estate (53\%), but lower than for healthcare (62\%) in the U.S.\footnote{According to the most recently available U.S. Economic Census data, available at \url{http://www.census.gov/econ/census/help/sector/data_topics/concentration_ratios.html}}

\subsection{Using PROWISH-HHI to determine thresholds for regulator intervention}
The analysis above is offered as an example of the kind of data that regulators could collect and use to inform their decisions about mergers and acquisitions in the third-party tracker market. Typically, such decisions are informed by the extent to which a particular instance of consolidation might raise the overall HHI for an industry, where market share is based on revenue or unit sales. We believe that such traditional measures should continue to form the primary basis for regulator action, and currently present a more extreme picture of market concentration; recent estimates suggest that Google and Facebook account for an estimated 72\% of digital ad spending in the U.S. and 54\% in the UK, whereas those firms account for just 31\% of the coverage in our combined web and mobile analysis ~\cite{slefo2017inapp,bold2017google}.

However, as suggested by EU policy-makers, competition regulators are in need of an equivalent measure for assessing a firm's share of total tracking capability. Grounded in systematic, empirical data analysis, our measures are designed to provide a more suitable solution for such needs. As mentioned above, privacy may be affected by tracker market concentration, but the implications of high concentration could be positive or negative depending on the different privacy concerns of affected individuals.

Neither the mobile nor the web tracking ecosystems are sufficiently concentrated, according to our PROWISH-HHI, to meet the threshold for intervention under U.S. competition law (0.25). However, both exceed the threshold used by E.U. regulators to determine when intervention may be required (0.1). As noted above, E.U. competition authorities generally consider any consolidation which is likely to raise the overall HHI of an industry by 0.025 (and above 0.1 overall), to be grounds for investigation ~\cite{verouden2004role}. Only one of the parent-subsidiary pairs in our records met this criteria (Google - Doubleclick). If Doubleclick were not owned by Google, the market concentration of web trackers would be 0.039 lower (according to our PROWISH-HHI measure) than it is at present. We also note that there are multiple pairs of trackers in the top 10 whose combined share would raise the HHI beyond the 0.025 threshold (e.g. Twitter and Yahoo).\footnote{This is example is merely illustrative, although such a merger was discussed according to some media reports ~\cite{Covert2016}}

\subsection{Limitations and future work}

Our approach has some noteworthy limitations, and there are a number of opportunities to refine it in future work.

Our approach to crawling websites and applications might present some confounding factors. Since our web crawling method maintains the browser state, the order in which sites are visited might affect what third-party content will be served. The possible effect of statefulness (and visitation order) is therefore unknown. We also only crawled the first page of a site, and similarly, during dynamic analysis of apps, we only ran the app for 30 seconds. Further use of these web pages or apps may have resulted in the detection of more trackers (although initial tests suggested diminishing returns on subsequent use).

We also made a number of assumptions about the relationships between parents and subsidiaries, and between organisations with no ownership relation. We ignored commitments made by parent firms not to merge the tracking capability of their subsidiaries with their own existing tracking infrastructure. If these promises take the form of sufficiently robust and binding agreements, increases in a firm's tracker share might pose less of a risk. However, recent cases suggest that promises made by parent companies during acquisitions may easily later be broken.\footnote{See e.g. ~\cite{whatsapp2016share,google2016broken}.} Similarly, we did not take into account the possibility that two or more tracker networks might enter into a business agreement to share data with one another. It may therefore be useful to incorporate the extent of cross-network data sharing, possibly through detection of cookie-syncing (as in e.g. ~\cite{englehardt2016census}). This is left for future work.

Another limitation of our approach is that the existence of counter-transparency efforts on the part of trackers might have reduced the completeness of our data. For instance, some of the hosts we identified in the web tracker analysis had domain names registered under proxy services, so the companies behind them could not be traced.\footnote{For further analysis of the use of such proxy services, see ~\cite{clayton2014study}} Similarly, about 3\% of the third party libraries we analysed in detail had been obfuscated to prevent reverse-engineering of application code.\footnote{Although various techniques have been proposed to circumvent such obfuscation using machine learning, e.g. ~\cite{ma2016libradar}.}  Further investigation of the strategies used by first-party developers and third-party network operators could provide further insight into this otherwise opaque corner of the tracking ecosystem.

The use of prominence and prevalence metrics also raises some important caveats. First, they do not take into account the sensitivity of the data captured by a tracker. Second, our prominence metric is based on the assumption that popularity follows a power law, so rank is roughly proportional to a website's audience. Actual number of users might make a better measure, but unfortunately this data was not available. Furthermore, prominence discounts trackers that are widely distributed on less popular sites / apps; but these less popular sites / apps might be considered as revealing more sensitive information about a user because they are niche.

It is also worth noting that tracking capability can only ever form part of an overall analysis of a firm's market power. In digital markets, dominance can be leveraged in multiple ways, including through `tieing' and `bundling' (refusing to sell one product unless the buyer also takes another)~\cite{edelman2015does}. As such, regulators must consider the tracking market in relation to a firm's other business activities, including provision of other services, hardware or software that relate to its tracking operations. For instance, Google not only operates the most prominent third-party tracker network in the Android mobile ecosystem, it also controls the de facto Android application marketplace, develops the operating system and its standard applications; these activities cannot be considered in isolation ~\cite{edelman2016android}. The ways in which a firm's dominance in tracking might spill over into other spheres of business are therefore important to consider.

Other avenues for future work include examining potential geographical differences in tracker coverage, to investigate the possible effect of different regulatory environments on tracking behaviour (see e.g. ~\cite{fruchter2015variations}). Exploring the trade-offs and incentives faced by app and website developers when they choose particular third-party trackers may also prove instructive in understanding how certain firms come to dominate the market.


\subsection{Conclusion}

As noted in the introduction, there are widespread concerns that too much power is falling into the hands of a small number of companies, who have the ability to track users across multiple websites and applications. Such concerns have the potential to be at least partially addressed from within existing competition and antitrust policy frameworks, but only if we have a systematic, empirically-grounded and principled means of quantifying the relative share of tracking capability of each company. We believe that the methods and metrics outlined above provide a practical, meaningful and fair approach for such analysis.

\section*{APPENDIX}
\setcounter{section}{1}
In order to evaluate the various possible methods of identifying potential trackers amongst websites and apps, we conducted several experimental tests on a subset of 200 apps and websites. Regarding mobile applications, we found that dynamic and static methods recalled different trackers (see table VI). On average, dynamic methods recalled 2.26 (subsidiary) / 2.02 (parent) trackers which the static methods failed to detect. Static methods had a better recall, with an average of 4.85 (subsidiary) / 3.45 (parent) trackers which the dynamic methods failed to detect. The average size of the intersection of mobile and web trackers found by each method was 2.9.

Second, we tested for disparities in the overlap between web and mobile trackers when using static or dynamic methods to detect mobile trackers (see table VII). We found that the highest overlap rate was obtained when using static methods for mobile and dynamic methods for the web, which is the combination of methods used in our main analysis of the 5,000 apps and 5,000 websites.

There are various possible explanations for these differences in recall. Dynamic methods may fail to detect certain trackers because they only appear in network traffic in response to user interactions not simulated during the session. Static methods may fail to detect certain trackers because URLs may not be referenced within the APK bytecode, but instead retrieved from a server at runtime, then used in subsequent requests. Another reason might be obfuscation. Popular obfuscation tools like Proguard only obfuscate classes and attributes, so one should expect any URL strings to be detected. However, such URLs may still be encoded or encrypted in the APK, and only decoded / decrypted at runtime (tools like Dexguard or Stringer Java Obfuscator enable this). Or they could be bit-shifted, or stored as character arrays and later converted to strings. These obfuscation techniques are all breakable in principle, but will be idiosyncratic to each developer so corresponding de-obfuscation techniques will also have to be suited to the particular methods used by the app in question.

\begin{table}[!htb]
\caption{Dynamic vs. static mobile tracker detection recall}{
\centering
\label{dyn-vs-static-mobile}
\begin{tabular}{|P{1in}|P{1in}|P{1in}|P{1in}|}
\hline
\textbf{Organisation level} & \textbf{Dynamic - Static} & \textbf{Static - Dynamic} & \textbf{Static $\cap$ Dynamic} \\ \hline
Subsidiary                  & 2.26                      & 4.85                      & 2.9                       \\ \hline
Parent                      & 2.02                      & 3.45                      & 2.9                       \\ \hline
\end{tabular}
}
\end{table}

\begin{table}[!htb]
\caption{Web vs. mobile tracker detection (dynamic vs static)}{
\centering
\label{dyn-web-vs-static-mobile}
\begin{tabular}{|P{0.75in}|P{0.75in}|P{1in}|P{0.5in}|P{0.5in}|P{0.5in}|}
\hline
\textbf{Mobile method}   & \textbf{Web method}      & \textbf{Organisation level} & \textbf{Mobile - Web} & \textbf{Web - Mobile} & \textbf{Mobile $\cap$ Web} \\ \hline
Dynamic &  Dynamic & Subsidiary                  & 3.4                  & 4.5                  & 1.63                   \\ \cline{3-6} 
                         &                          & Parent                      & 3.17                  & 4.36                  & 1.63                   \\ \hline
Static & Dynamic & Subsidiary                  & 5.97                  & 4.6                   & 1.63                  \\ \cline{3-6} 
                         &                          & Parent                      & 4.24                  & 4.06                  & 2.0                   \\ \hline
\end{tabular}
}
\end{table}

\begin{acks}
All authors were supported under \emph{SOCIAM: The Theory and Practice of Social Machines.} The SOCIAM Project is funded by the UK Engineering and Physical Sciences Research Council (EPSRC) under grant number EP/J017728/2 and comprises the University of Oxford, the University of Southampton, and the University of Edinburgh. The authors would also like to thank Vincent Taylor of Magdalen College, University of Oxford, for his help in procuring APKs, and Dr Greg Taylor of the Oxford Internet Institute, for his advice on the application of economic indicators in online marketplaces.
\end{acks}

\bibliographystyle{ACM-Reference-Format}
\bibliography{sample-bibliography}


\begin{thebibliography}{47}


\ifx \showCODEN    \undefined \def \showCODEN     #1{\unskip}     \fi
\ifx \showDOI      \undefined \def \showDOI       #1{#1}\fi
\ifx \showISBNx    \undefined \def \showISBNx     #1{\unskip}     \fi
\ifx \showISBNxiii \undefined \def \showISBNxiii  #1{\unskip}     \fi
\ifx \showISSN     \undefined \def \showISSN      #1{\unskip}     \fi
\ifx \showLCCN     \undefined \def \showLCCN      #1{\unskip}     \fi
\ifx \shownote     \undefined \def \shownote      #1{#1}          \fi
\ifx \showarticletitle \undefined \def \showarticletitle #1{#1}   \fi
\ifx \showURL      \undefined \def \showURL       {\relax}        \fi
\providecommand\bibfield[2]{#2}
\providecommand\bibinfo[2]{#2}
\providecommand\natexlab[1]{#1}
\providecommand\showeprint[2][]{arXiv:#2}

\bibitem[\protect\citeauthoryear{Acar, Juarez, Nikiforakis, Diaz, G{\"u}rses,
  Piessens, and Preneel}{Acar et~al\mbox{.}}{2013}]%
        {acar2013fpdetective}
\bibfield{author}{\bibinfo{person}{Gunes Acar}, \bibinfo{person}{Marc Juarez},
  \bibinfo{person}{Nick Nikiforakis}, \bibinfo{person}{Claudia Diaz},
  \bibinfo{person}{Seda G{\"u}rses}, \bibinfo{person}{Frank Piessens}, {and}
  \bibinfo{person}{Bart Preneel}.} \bibinfo{year}{2013}\natexlab{}.
\newblock \showarticletitle{FPDetective: dusting the web for fingerprinters}.
  In \bibinfo{booktitle}{{\em Proc. of ACM SIGSAC conference on Computer \&
  communications security}}. ACM, \bibinfo{pages}{1129--1140}.
\newblock


\bibitem[\protect\citeauthoryear{Acquisti, Taylor, and Wagman}{Acquisti
  et~al\mbox{.}}{2016}]%
        {acquisti2016economics}
\bibfield{author}{\bibinfo{person}{Alessandro Acquisti},
  \bibinfo{person}{Curtis~R Taylor}, {and} \bibinfo{person}{Liad Wagman}.}
  \bibinfo{year}{2016}\natexlab{}.
\newblock \showarticletitle{The economics of privacy}.
\newblock \bibinfo{journal}{{\em Journal of Economic Literature\/}}
  \bibinfo{volume}{52}, \bibinfo{number}{2} (\bibinfo{year}{2016}).
\newblock


\bibitem[\protect\citeauthoryear{Anderson, Bonneau, and Stajano}{Anderson
  et~al\mbox{.}}{2010}]%
        {anderson2010inglorious}
\bibfield{author}{\bibinfo{person}{Jonathan Anderson}, \bibinfo{person}{Joseph
  Bonneau}, {and} \bibinfo{person}{Frank Stajano}.}
  \bibinfo{year}{2010}\natexlab{}.
\newblock \showarticletitle{Inglorious Installers: Security in the Application
  Marketplace.}. In \bibinfo{booktitle}{{\em WEIS}}. Citeseer.
\newblock


\bibitem[\protect\citeauthoryear{Angwin}{Angwin}{2016}]%
        {google2016broken}
\bibfield{author}{\bibinfo{person}{Julia Angwin}.}
  \bibinfo{year}{2016}\natexlab{}.
\newblock \showarticletitle{Google's Broken Privacy Promise}.
\newblock \bibinfo{journal}{{\em Pacific Standard Magazine\/}}
  (\bibinfo{year}{2016}).
\newblock


\bibitem[\protect\citeauthoryear{Arzt, Rasthofer, Fritz, Bodden, Bartel, Klein,
  Le~Traon, Octeau, and McDaniel}{Arzt et~al\mbox{.}}{2014}]%
        {arzt2014flowdroid}
\bibfield{author}{\bibinfo{person}{Steven Arzt}, \bibinfo{person}{Siegfried
  Rasthofer}, \bibinfo{person}{Christian Fritz}, \bibinfo{person}{Eric Bodden},
  \bibinfo{person}{Alexandre Bartel}, \bibinfo{person}{Jacques Klein},
  \bibinfo{person}{Yves Le~Traon}, \bibinfo{person}{Damien Octeau}, {and}
  \bibinfo{person}{Patrick McDaniel}.} \bibinfo{year}{2014}\natexlab{}.
\newblock \showarticletitle{Flowdroid: Precise context, flow, field,
  object-sensitive and lifecycle-aware taint analysis for android apps}.
\newblock \bibinfo{journal}{{\em ACM SIGPLAN Notices\/}} \bibinfo{volume}{49},
  \bibinfo{number}{6} (\bibinfo{year}{2014}), \bibinfo{pages}{259--269}.
\newblock


\bibitem[\protect\citeauthoryear{Aziz and Telang}{Aziz and Telang}{2015}]%
        {aziz2015cookie}
\bibfield{author}{\bibinfo{person}{Arslan Aziz} {and} \bibinfo{person}{Rahul
  Telang}.} \bibinfo{year}{2015}\natexlab{}.
\newblock \bibinfo{booktitle}{{\em What is a Cookie Worth?}}
\newblock \bibinfo{type}{{T}echnical {R}eport}. \bibinfo{institution}{Technical
  Report}.
\newblock


\bibitem[\protect\citeauthoryear{Bailey}{Bailey}{2016}]%
        {whatsapp2016share}
\bibfield{author}{\bibinfo{person}{Brandon Bailey}.}
  \bibinfo{year}{2016}\natexlab{}.
\newblock \showarticletitle{WhatsApp is going to share your phone number with
  Facebook}.
\newblock \bibinfo{journal}{{\em Big Story, Associated Press\/}}
  (\bibinfo{year}{2016}).
\newblock


\bibitem[\protect\citeauthoryear{Batyuk, Herpich, Camtepe, Raddatz, Schmidt,
  and Albayrak}{Batyuk et~al\mbox{.}}{2011}]%
        {batyuk2011using}
\bibfield{author}{\bibinfo{person}{Leonid Batyuk}, \bibinfo{person}{Markus
  Herpich}, \bibinfo{person}{Seyit~Ahmet Camtepe}, \bibinfo{person}{Karsten
  Raddatz}, \bibinfo{person}{Aubrey-Derrick Schmidt}, {and}
  \bibinfo{person}{Sahin Albayrak}.} \bibinfo{year}{2011}\natexlab{}.
\newblock \showarticletitle{Using static analysis for automatic assessment and
  mitigation of unwanted and malicious activities within Android applications}.
  In \bibinfo{booktitle}{{\em Malicious and Unwanted Software (MALWARE), 2011
  6th International Conference on}}. IEEE, \bibinfo{pages}{66--72}.
\newblock


\bibitem[\protect\citeauthoryear{Bold}{Bold}{2017}]%
        {bold2017google}
\bibfield{author}{\bibinfo{person}{Ben Bold}.} \bibinfo{year}{2017}\natexlab{}.
\newblock \showarticletitle{Google and Facebook dominate over half of digital
  media market}.
\newblock  (\bibinfo{year}{2017}).
\newblock
\showURL{%
\url{https://www.campaignlive.co.uk/article/google-facebook-dominate-half-digital-media-market/1444793}}


\bibitem[\protect\citeauthoryear{Book and Wallach}{Book and Wallach}{2015}]%
        {book2015empirical}
\bibfield{author}{\bibinfo{person}{Theodore Book} {and} \bibinfo{person}{Dan~S
  Wallach}.} \bibinfo{year}{2015}\natexlab{}.
\newblock \showarticletitle{An empirical study of mobile ad targeting}.
\newblock \bibinfo{journal}{{\em arXiv preprint arXiv:1502.06577\/}}
  (\bibinfo{year}{2015}).
\newblock


\bibitem[\protect\citeauthoryear{Clayton and Mansfield}{Clayton and
  Mansfield}{2014}]%
        {clayton2014study}
\bibfield{author}{\bibinfo{person}{Richard Clayton} {and} \bibinfo{person}{Tony
  Mansfield}.} \bibinfo{year}{2014}\natexlab{}.
\newblock \showarticletitle{A study of Whois privacy and proxy service abuse}.
  In \bibinfo{booktitle}{{\em Proceedings (online) of the 13th Workshop on
  Economics of Information Security, State College, PA (June 2014)}}.
\newblock


\bibitem[\protect\citeauthoryear{Cline}{Cline}{2013}]%
        {geradin2013}
\bibfield{author}{\bibinfo{person}{Jay Cline}.}
  \bibinfo{year}{2013}\natexlab{}.
\newblock \showarticletitle{U.S. takes the gold in doling out privacy fines}.
\newblock \bibinfo{journal}{{\em Available at SSRN 2216088\/}}
  (\bibinfo{year}{2013}).
\newblock


\bibitem[\protect\citeauthoryear{Covert}{Covert}{2016}]%
        {Covert2016}
\bibfield{author}{\bibinfo{person}{James Covert}.}
  \bibinfo{year}{2016}\natexlab{}.
\newblock \showarticletitle{Twitter kicked tires on Yahoo merger}.
\newblock  (\bibinfo{year}{2016}).
\newblock
\showURL{%
\url{http://nypost.com/2016/06/02/twitter-talks-to-yahoo-about-merger/}}


\bibitem[\protect\citeauthoryear{Edelman}{Edelman}{2015}]%
        {edelman2015does}
\bibfield{author}{\bibinfo{person}{Benjamin Edelman}.}
  \bibinfo{year}{2015}\natexlab{}.
\newblock \showarticletitle{Does Google leverage market power through tying and
  bundling?}
\newblock \bibinfo{journal}{{\em Journal of Competition Law and Economics\/}}
  \bibinfo{volume}{11}, \bibinfo{number}{2} (\bibinfo{year}{2015}),
  \bibinfo{pages}{365--400}.
\newblock


\bibitem[\protect\citeauthoryear{Edelman and Geradin}{Edelman and
  Geradin}{2016}]%
        {edelman2016android}
\bibfield{author}{\bibinfo{person}{Benjamin~G Edelman} {and}
  \bibinfo{person}{Damien Geradin}.} \bibinfo{year}{2016}\natexlab{}.
\newblock \showarticletitle{Android and Competition Law: Exploring and
  Assessing Google's Practices in Mobile}.
\newblock  (\bibinfo{year}{2016}).
\newblock


\bibitem[\protect\citeauthoryear{Egele, Kruegel, Kirda, and Vigna}{Egele
  et~al\mbox{.}}{2011}]%
        {egele2011pios}
\bibfield{author}{\bibinfo{person}{Manuel Egele}, \bibinfo{person}{Christopher
  Kruegel}, \bibinfo{person}{Engin Kirda}, {and} \bibinfo{person}{Giovanni
  Vigna}.} \bibinfo{year}{2011}\natexlab{}.
\newblock \showarticletitle{PiOS: Detecting Privacy Leaks in iOS
  Applications.}. In \bibinfo{booktitle}{{\em NDSS}}.
  \bibinfo{pages}{177--183}.
\newblock


\bibitem[\protect\citeauthoryear{Enck, Gilbert, Han, Tendulkar, Chun, Cox,
  Jung, McDaniel, and Sheth}{Enck et~al\mbox{.}}{2014}]%
        {enck2014taintdroid}
\bibfield{author}{\bibinfo{person}{William Enck}, \bibinfo{person}{Peter
  Gilbert}, \bibinfo{person}{Seungyeop Han}, \bibinfo{person}{Vasant
  Tendulkar}, \bibinfo{person}{Byung-Gon Chun}, \bibinfo{person}{Landon~P Cox},
  \bibinfo{person}{Jaeyeon Jung}, \bibinfo{person}{Patrick McDaniel}, {and}
  \bibinfo{person}{Anmol~N Sheth}.} \bibinfo{year}{2014}\natexlab{}.
\newblock \showarticletitle{TaintDroid: an information-flow tracking system for
  realtime privacy monitoring on smartphones}.
\newblock \bibinfo{journal}{{\em ACM Transactions on Computer Systems
  (TOCS)\/}} \bibinfo{volume}{32}, \bibinfo{number}{2} (\bibinfo{year}{2014}),
  \bibinfo{pages}{5}.
\newblock


\bibitem[\protect\citeauthoryear{Englehardt and Narayanan}{Englehardt and
  Narayanan}{2016}]%
        {englehardt2016census}
\bibfield{author}{\bibinfo{person}{Steven Englehardt} {and}
  \bibinfo{person}{Arvind Narayanan}.} \bibinfo{year}{2016}\natexlab{}.
\newblock \showarticletitle{{Online tracking: A 1-million-site measurement and
  analysis}}. In \bibinfo{booktitle}{{\em Proceedings of ACM Computer and
  Communications Security 2016}}.
\newblock


\bibitem[\protect\citeauthoryear{Englehardt, Reisman, Eubank, Zimmerman, Mayer,
  Narayanan, and Felten}{Englehardt et~al\mbox{.}}{2015}]%
        {englehardt2015cookies}
\bibfield{author}{\bibinfo{person}{Steven Englehardt}, \bibinfo{person}{Dillon
  Reisman}, \bibinfo{person}{Christian Eubank}, \bibinfo{person}{Peter
  Zimmerman}, \bibinfo{person}{Jonathan Mayer}, \bibinfo{person}{Arvind
  Narayanan}, {and} \bibinfo{person}{Edward~W Felten}.}
  \bibinfo{year}{2015}\natexlab{}.
\newblock \showarticletitle{Cookies that give you away: The surveillance
  implications of web tracking}. In \bibinfo{booktitle}{{\em Proc. of the 24th
  International Conference on World Wide Web}}. ACM, \bibinfo{pages}{289--299}.
\newblock


\bibitem[\protect\citeauthoryear{Fruchter, Miao, Stevenson, and
  Balebako}{Fruchter et~al\mbox{.}}{2015}]%
        {fruchter2015variations}
\bibfield{author}{\bibinfo{person}{Nathaniel Fruchter}, \bibinfo{person}{Hsin
  Miao}, \bibinfo{person}{Scott Stevenson}, {and} \bibinfo{person}{Rebecca
  Balebako}.} \bibinfo{year}{2015}\natexlab{}.
\newblock \showarticletitle{Variations in tracking in relation to geographic
  location}.
\newblock \bibinfo{journal}{{\em arXiv preprint arXiv:1506.04103\/}}
  (\bibinfo{year}{2015}).
\newblock


\bibitem[\protect\citeauthoryear{Harbour}{Harbour}{2007}]%
        {harbourdissenting}
\bibfield{author}{\bibinfo{person}{Commissioner Pamela~Jones Harbour}.}
  \bibinfo{year}{2007}\natexlab{}.
\newblock \showarticletitle{Dissenting Statement of Commissioner Pamela Jones
  Harbour and Commissioner J. Thomas Rosch in the Matter of Western Refining,
  Inc. et al., Docket No. 9323/ File No. 061-0259}.
\newblock  (\bibinfo{year}{2007}).
\newblock


\bibitem[\protect\citeauthoryear{Hustinx}{Hustinx}{2014}]%
        {hustinx2014privacy}
\bibfield{author}{\bibinfo{person}{P Hustinx}.}
  \bibinfo{year}{2014}\natexlab{}.
\newblock \bibinfo{title}{Privacy and Competitiveness in the Age of Big Data:
  Preliminary Opinion of the European Data Protection Supervisor}.
\newblock   (\bibinfo{year}{2014}).
\newblock


\bibitem[\protect\citeauthoryear{IAB}{IAB}{2016}]%
        {iab2016}
\bibfield{author}{\bibinfo{person}{IAB}.} \bibinfo{year}{2016}\natexlab{}.
\newblock \bibinfo{title}{IAB Internet Advertising Revenue Report 2015}.
\newblock   (\bibinfo{year}{2016}).
\newblock


\bibitem[\protect\citeauthoryear{Lande}{Lande}{2008}]%
        {lande2008microsoft}
\bibfield{author}{\bibinfo{person}{Robert~H Lande}.}
  \bibinfo{year}{2008}\natexlab{}.
\newblock \showarticletitle{The Microsoft-Yahoo Merger: Yes, Privacy is an
  Antitrust Concern}.
\newblock \bibinfo{journal}{{\em FTC: Watch\/}} \bibinfo{number}{714}
  (\bibinfo{year}{2008}).
\newblock


\bibitem[\protect\citeauthoryear{Leung, Ren, Choffnes, and Wilson}{Leung
  et~al\mbox{.}}{2016}]%
        {leung2016recon}
\bibfield{author}{\bibinfo{person}{Christophe Leung}, \bibinfo{person}{Jingjing
  Ren}, \bibinfo{person}{David Choffnes}, {and} \bibinfo{person}{Christo
  Wilson}.} \bibinfo{year}{2016}\natexlab{}.
\newblock \showarticletitle{Should You Use the App for That? Comparing the
  Privacy Implications of App-and Web-based Online Services}. In
  \bibinfo{booktitle}{{\em Proc. of the 16th ACM Internet Measurement
  Conference}}. \bibinfo{pages}{To appear}.
\newblock


\bibitem[\protect\citeauthoryear{Libert}{Libert}{2015}]%
        {libert2015exposing}
\bibfield{author}{\bibinfo{person}{Timothy Libert}.}
  \bibinfo{year}{2015}\natexlab{}.
\newblock \showarticletitle{Exposing the Invisible Web: An Analysis of
  Third-Party HTTP Requests on 1 Million Websites}.
\newblock \bibinfo{journal}{{\em International Journal of Communication\/}}
  \bibinfo{volume}{9} (\bibinfo{year}{2015}), \bibinfo{pages}{18}.
\newblock


\bibitem[\protect\citeauthoryear{Lin, Liu, Sadeh, and Hong}{Lin
  et~al\mbox{.}}{2014}]%
        {lin2014privacygrade}
\bibfield{author}{\bibinfo{person}{Jialiu Lin}, \bibinfo{person}{Bin Liu},
  \bibinfo{person}{Norman Sadeh}, {and} \bibinfo{person}{Jason~I. Hong}.}
  \bibinfo{year}{2014}\natexlab{}.
\newblock \showarticletitle{Modeling Users{\textquoteright} Mobile App Privacy
  Preferences: Restoring Usability in a Sea of Permission Settings}. In
  \bibinfo{booktitle}{{\em Symposium On Usable Privacy and Security (SOUPS
  2014)}}. \bibinfo{publisher}{USENIX Association}, \bibinfo{address}{Menlo
  Park, CA}, \bibinfo{pages}{199--212}.
\newblock
\showISBNx{978-1-931971-13-3}
\showURL{%
\url{https://www.usenix.org/conference/soups2014/proceedings/presentation/lin}}


\bibitem[\protect\citeauthoryear{Ma, Wang, Guo, and Chen}{Ma
  et~al\mbox{.}}{2016}]%
        {ma2016libradar}
\bibfield{author}{\bibinfo{person}{Ziang Ma}, \bibinfo{person}{Haoyu Wang},
  \bibinfo{person}{Yao Guo}, {and} \bibinfo{person}{Xiangqun Chen}.}
  \bibinfo{year}{2016}\natexlab{}.
\newblock \showarticletitle{LibRadar: fast and accurate detection of
  third-party libraries in Android apps}. In \bibinfo{booktitle}{{\em
  Proceedings of the 38th International Conference on Software Engineering
  Companion}}. ACM, \bibinfo{pages}{653--656}.
\newblock


\bibitem[\protect\citeauthoryear{Meier}{Meier}{2014}]%
        {meier2014erfolgreicher}
\bibfield{author}{\bibinfo{person}{Saskia Meier}.}
  \bibinfo{year}{2014}\natexlab{}.
\newblock \bibinfo{booktitle}{{\em Erfolgreicher Anzeigenverkauf in mobilen
  Medien: Eine empirische Analyse zu Verkaufsindikatoren im Mobile
  Advertising}}.
\newblock \bibinfo{publisher}{Springer-Verlag}.
\newblock


\bibitem[\protect\citeauthoryear{Montes, Sand-Zantman, and Valletti}{Montes
  et~al\mbox{.}}{2015}]%
        {montes2015value}
\bibfield{author}{\bibinfo{person}{Rodrigo Montes}, \bibinfo{person}{Wilfried
  Sand-Zantman}, {and} \bibinfo{person}{Tommaso~M Valletti}.}
  \bibinfo{year}{2015}\natexlab{}.
\newblock \showarticletitle{The value of personal information in markets with
  endogenous privacy}.
\newblock  (\bibinfo{year}{2015}).
\newblock


\bibitem[\protect\citeauthoryear{OECD}{OECD}{2015}]%
        {oecd2015data}
\bibfield{author}{\bibinfo{person}{OECD}.} \bibinfo{year}{2015}\natexlab{}.
\newblock \showarticletitle{Data-Driven Innovation}.
\newblock  (\bibinfo{year}{2015}).
\newblock
\showDOI{%
\url{https://doi.org/10.1787/9789264229358-en}}


\bibitem[\protect\citeauthoryear{Olejnik, Minh-Dung, and Castelluccia}{Olejnik
  et~al\mbox{.}}{2013}]%
        {olejnik2013selling}
\bibfield{author}{\bibinfo{person}{Lukasz Olejnik}, \bibinfo{person}{Tran
  Minh-Dung}, {and} \bibinfo{person}{Claude Castelluccia}.}
  \bibinfo{year}{2013}\natexlab{}.
\newblock \showarticletitle{Selling off privacy at auction}.
\newblock  (\bibinfo{year}{2013}).
\newblock


\bibitem[\protect\citeauthoryear{Pasquale}{Pasquale}{2013}]%
        {pasquale2013privacy}
\bibfield{author}{\bibinfo{person}{Frank~A Pasquale}.}
  \bibinfo{year}{2013}\natexlab{}.
\newblock \showarticletitle{Privacy, Antitrust, and Power}.
\newblock \bibinfo{journal}{{\em George Mason Law Review\/}}
  \bibinfo{volume}{20}, \bibinfo{number}{4} (\bibinfo{year}{2013}),
  \bibinfo{pages}{1009--1024}.
\newblock


\bibitem[\protect\citeauthoryear{Qiu, Zhang, Shen, and Sun}{Qiu
  et~al\mbox{.}}{2015}]%
        {qiu2015apptrace}
\bibfield{author}{\bibinfo{person}{Lingzhi Qiu}, \bibinfo{person}{Zixiong
  Zhang}, \bibinfo{person}{Ziyi Shen}, {and} \bibinfo{person}{Guozi Sun}.}
  \bibinfo{year}{2015}\natexlab{}.
\newblock \showarticletitle{AppTrace: Dynamic trace on Android devices}. In
  \bibinfo{booktitle}{{\em 2015 IEEE International Conference on
  Communications}}. IEEE, \bibinfo{pages}{7145--7150}.
\newblock


\bibitem[\protect\citeauthoryear{Ren, Rao, Lindorfer, Legout, and Choffnes}{Ren
  et~al\mbox{.}}{2016}]%
        {ren2016recon}
\bibfield{author}{\bibinfo{person}{Jingjing Ren}, \bibinfo{person}{Ashwin Rao},
  \bibinfo{person}{Martina Lindorfer}, \bibinfo{person}{Arnaud Legout}, {and}
  \bibinfo{person}{David Choffnes}.} \bibinfo{year}{2016}\natexlab{}.
\newblock \showarticletitle{Demo: ReCon: Revealing and Controlling PII Leaks in
  Mobile Network Traffic}. In \bibinfo{booktitle}{{\em Proceedings of the
  International Conference on Mobile Systems, Applications, and Services
  Companion}} {\em (\bibinfo{series}{MobiSys '16 Companion})}.
  \bibinfo{pages}{117--117}.
\newblock


\bibitem[\protect\citeauthoryear{Roesner, Kohno, and Wetherall}{Roesner
  et~al\mbox{.}}{2012}]%
        {roesner2012detecting}
\bibfield{author}{\bibinfo{person}{Franziska Roesner},
  \bibinfo{person}{Tadayoshi Kohno}, {and} \bibinfo{person}{David Wetherall}.}
  \bibinfo{year}{2012}\natexlab{}.
\newblock \showarticletitle{Detecting and defending against third-party
  tracking on the web}. In \bibinfo{booktitle}{{\em Proc. of the 9th USENIX
  conference on Networked Systems Design and Implementation}}. USENIX
  Association, \bibinfo{pages}{12--12}.
\newblock


\bibitem[\protect\citeauthoryear{Schechner}{Schechner}{2017a}]%
        {schechner2017germany}
\bibfield{author}{\bibinfo{person}{Sam Schechner}.}
  \bibinfo{year}{2017}\natexlab{a}.
\newblock \showarticletitle{Germany Says Facebook Abuses Market Dominance to
  Collect Data}.
\newblock  (\bibinfo{year}{2017}).
\newblock
\showURL{%
\url{https://www.wsj.com/articles/facebook-abuses-its-dominance-to-harvest-your-data-says-german-antitrust-enforcer-1513680355}}


\bibitem[\protect\citeauthoryear{Schechner}{Schechner}{2017b}]%
        {slefo2017inapp}
\bibfield{author}{\bibinfo{person}{Sam Schechner}.}
  \bibinfo{year}{2017}\natexlab{b}.
\newblock \showarticletitle{In-App Mobile Ad Spend to reach 45.3bn, Facebook
  and Google Rejoice}.
\newblock  (\bibinfo{year}{2017}).
\newblock
\showURL{%
\url{http://adage.com/article/digital/facebook-google-dominate-45-billion-app-ad-market/310761/}}


\bibitem[\protect\citeauthoryear{Shapiro}{Shapiro}{2010}]%
        {shapiro20102010}
\bibfield{author}{\bibinfo{person}{Carl Shapiro}.}
  \bibinfo{year}{2010}\natexlab{}.
\newblock \showarticletitle{The 2010 horizontal merger guidelines: From
  hedgehog to fox in forty years}.
\newblock \bibinfo{journal}{{\em Antitrust Law Journal\/}}
  \bibinfo{volume}{77}, \bibinfo{number}{1} (\bibinfo{year}{2010}),
  \bibinfo{pages}{49--107}.
\newblock


\bibitem[\protect\citeauthoryear{Swire}{Swire}{2007}]%
        {swire2007protecting}
\bibfield{author}{\bibinfo{person}{Peter Swire}.}
  \bibinfo{year}{2007}\natexlab{}.
\newblock \bibinfo{title}{Protecting Consumers: Privacy Matters in Antitrust
  Analysis}.
\newblock   (\bibinfo{year}{2007}).
\newblock


\bibitem[\protect\citeauthoryear{Taylor and Martinovic}{Taylor and
  Martinovic}{2017}]%
        {taylor2017}
\bibfield{author}{\bibinfo{person}{V.~F. Taylor} {and} \bibinfo{person}{I.
  Martinovic}.} \bibinfo{year}{2017}\natexlab{}.
\newblock \showarticletitle{To Update or Not to Update: Insights From a
  Two-Year Study of Android App Evolution}. In \bibinfo{booktitle}{{\em ACM
  Asia Conference on Computer and Communications Security (ASIACCS'17)}}.
\newblock
\showDOI{%
\url{https://doi.org/10}}


\bibitem[\protect\citeauthoryear{Vallina-Rodriguez, Sundaresan, Razaghpanah,
  Nithyanand, Allman, Kreibich, and Gill}{Vallina-Rodriguez
  et~al\mbox{.}}{2016}]%
        {vallina2016tracking}
\bibfield{author}{\bibinfo{person}{Narseo Vallina-Rodriguez},
  \bibinfo{person}{Srikanth Sundaresan}, \bibinfo{person}{Abbas Razaghpanah},
  \bibinfo{person}{Rishab Nithyanand}, \bibinfo{person}{Mark Allman},
  \bibinfo{person}{Christian Kreibich}, {and} \bibinfo{person}{Phillipa Gill}.}
  \bibinfo{year}{2016}\natexlab{}.
\newblock \showarticletitle{Tracking the Trackers: Towards Understanding the
  Mobile Advertising and Tracking Ecosystem}.
\newblock \bibinfo{journal}{{\em arXiv preprint arXiv:1609.07190\/}}
  (\bibinfo{year}{2016}).
\newblock


\bibitem[\protect\citeauthoryear{Van~Kleek, Liccardi, Binns, Zhao, Weitzner,
  and Shadbolt}{Van~Kleek et~al\mbox{.}}{2017}]%
        {van2017better}
\bibfield{author}{\bibinfo{person}{Max Van~Kleek}, \bibinfo{person}{Ilaria
  Liccardi}, \bibinfo{person}{Reuben Binns}, \bibinfo{person}{Jun Zhao},
  \bibinfo{person}{Daniel~J Weitzner}, {and} \bibinfo{person}{Nigel Shadbolt}.}
  \bibinfo{year}{2017}\natexlab{}.
\newblock \showarticletitle{Better the devil you know: Exposing the data
  sharing practices of smartphone apps}. In \bibinfo{booktitle}{{\em
  Proceedings of the 2017 CHI Conference on Human Factors in Computing
  Systems}}. ACM, \bibinfo{pages}{5208--5220}.
\newblock


\bibitem[\protect\citeauthoryear{Verouden}{Verouden}{2004}]%
        {verouden2004role}
\bibfield{author}{\bibinfo{person}{Vincent Verouden}.}
  \bibinfo{year}{2004}\natexlab{}.
\newblock \showarticletitle{The role of market shares and market concentration
  indices in the European Commission's Guidelines on the assessment of
  horizontal mergers under the EC Merger Regulation}. In
  \bibinfo{booktitle}{{\em Comments prepared for the FTC and US DOJ Merger
  Enforcement Workshop, Washington, DC}}.
\newblock


\bibitem[\protect\citeauthoryear{Warren-Boulton}{Warren-Boulton}{1990}]%
        {warren1990implications}
\bibfield{author}{\bibinfo{person}{Frederick~R Warren-Boulton}.}
  \bibinfo{year}{1990}\natexlab{}.
\newblock \showarticletitle{Implications of US experience with horizontal
  mergers and takeovers for Canadian competition policy}.
\newblock \bibinfo{journal}{{\em The law and economics of competition policy.
  Vancouver, BC\/}} (\bibinfo{year}{1990}).
\newblock


\bibitem[\protect\citeauthoryear{Yu, Macbeth, Modi, and Pujol}{Yu
  et~al\mbox{.}}{2016}]%
        {yu2016tracking}
\bibfield{author}{\bibinfo{person}{Zhonghao Yu}, \bibinfo{person}{Sam Macbeth},
  \bibinfo{person}{Konark Modi}, {and} \bibinfo{person}{Josep~M Pujol}.}
  \bibinfo{year}{2016}\natexlab{}.
\newblock \showarticletitle{Tracking the Trackers}. In \bibinfo{booktitle}{{\em
  Proceedings of the 25th International Conference on World Wide Web}}.
  International World Wide Web Conferences Steering Committee,
  \bibinfo{pages}{121--132}.
\newblock


\bibitem[\protect\citeauthoryear{Zuckerman}{Zuckerman}{2014}]%
        {zuckerman2014internet}
\bibfield{author}{\bibinfo{person}{Ethan Zuckerman}.}
  \bibinfo{year}{2014}\natexlab{}.
\newblock \showarticletitle{The Internet's original sin}.
\newblock \bibinfo{journal}{{\em The Atlantic\/}}  \bibinfo{volume}{14}
  (\bibinfo{year}{2014}), \bibinfo{pages}{1--8}.
\newblock


\end{thebibliography}

\end{document}